\documentclass[slac_one]{revtex4}
\usepackage{graphicx}
\usepackage{fancyhdr}
\usepackage{amsmath}
\usepackage{subfigure}
\begin{document}

\title{{\small{*}}\\ 
\vspace{12pt}
Kinematical Analysis of an Articulated Mechanism} 

%

\author{L. Fleischfresser}
\affiliation{UTFPR, Campo Mour\~{a}o, PR 87301-899, BRAZIL}
%

\begin{abstract}
The purpose of this work is twofold: to present mathematical expressions for the kinematics of an articulated mechanism and to perform numerical experiments with the implemented code. The system of rigid parts is made of two slender bars and a disk. In the original configuration, a constant counterclockwise rotation rate is imposed on the disk. In the modified version, this angular velocity varies linearly with the rotation angle to produce an \emph{average} rate that is nearly the same as the constant case. Angles, velocities and accelerations are analyzed for a $90^{o}$ turn of the disk. The numerical solutions show the inversion of the linking bar sense of rotation along with the start of deceleration for both bars. The paper and pencil solution of the original problem that may lead to a wrong conclusion is explained. Equations are derived from first principles and the code is placed under version control. Those in charge of vector dynamics courses may find it useful as a project-based learning activity.
\end{abstract}

\maketitle

\thispagestyle{fancy}


\section{INTRODUCTION} 
The rigid body is an important idealization of moving and interacting parts and beings. The key feature is the fixed distance between two points in the body, since forces and torques cannot cause deformation of its shape~\citep{Stewart:2000}. The subject is relevant for the development of video-game physics engines since the simulation of motions and interactions of rigid bodies approximates reality fairly well~\citep{Garstenauer:2006}. Prosthetic limbs and robotic devices are other important application areas of this theme.

Here we offer a tool to stimulate engineering students to continue their learning path in the subject. We show mathematical formulations and numerical simulations of the interdependent motions.

\section{MATHEMATICAL FORMULATIONS}

\begin{figure*}[t]
\centering
\subfigure[Initial position]{
    \includegraphics[width=0.4\textwidth]{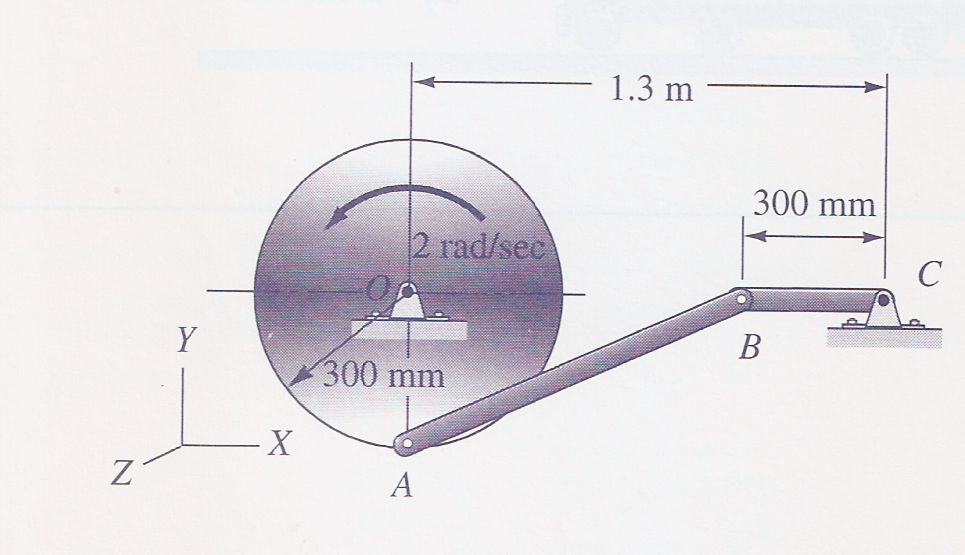}
    \label{fig:subfig1}
}
\subfigure[Intermediate position]{
    \includegraphics[width=0.4\textwidth]{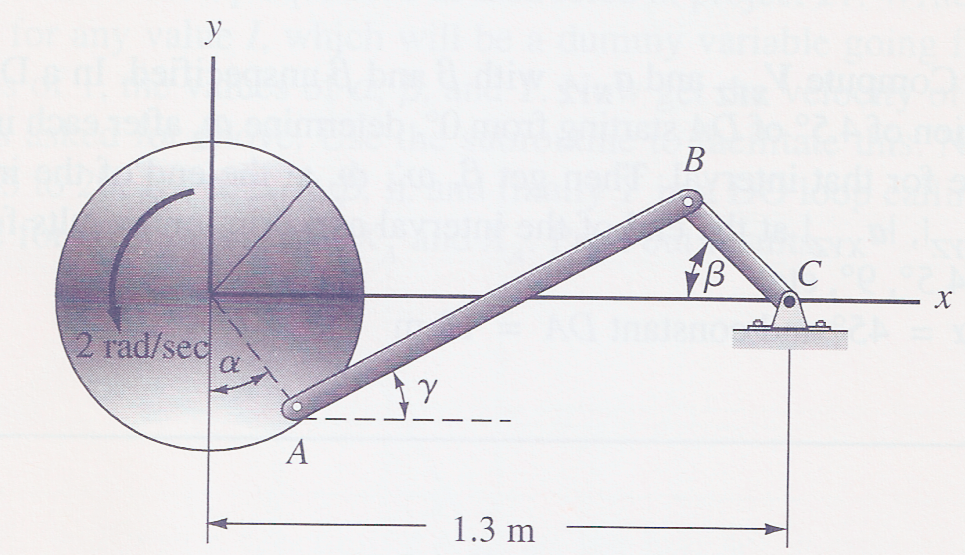}
    \label{fig:subfig2}
}
\caption[Optional caption for list of figures]{Articulated mechanism at its \subref{fig:subfig1} initial and \subref{fig:subfig2} intermediate positions. Reproduced from~\cite{Shames:1996}.}
\label{fig:both}
\end{figure*}

The analysis has its origins in~\cite{Shames:1996}, where there is a solved example for the instantaneous position as shown in~Fig.~\ref{fig:subfig1}, and general guidelines for a computer project are proposed. In what follows, equations for the kinematics are derived showing angle and length constraints, as well as velocity and acceleration relations while the disk undergoes either a constant $2~rad/s$, or a variable $\left( 0.10~+~2.3\cdot\,\alpha \right)\;rad/s$, with $\alpha$ being the turn angle of the disk~[Fig.~\ref{fig:subfig2}].

\subsection{Angles and length constraints}

The three angles describing rotation of the articulated mechanism are $\alpha$, $\beta$, and $\gamma$. A one letter notation is introduced to write trigonometric functions as:

\begin{align}
      l&=sin\, \alpha&\ \ \ \ m&=cos\, \alpha \\
      n&=sin\, \beta&\ \ \ \ o&=cos\, \beta \\
      p&=sin\, \gamma&\ \ \ \ q&=cos\, \gamma
\end{align}

Since the length of the linking bar AB cannot change, the horizontal displacement of pin A must be equal to the horizontal displacement of pin B, or:

\begin{equation}
\overline{OA} \cdot\; l\; =\; \overline{BC}\; -\; \overline{BC} \cdot\; o
\label{eq1}
\end{equation}

An overbar means the length of a given straight line segment. It can be inferred from Fig.~\ref{fig:subfig1} that $\overline{OA}\,=\,\overline{BC}$. Simplifying:

\begin{equation}
l\; =\; 1\; -\; o 
\label{eq2}
\end{equation}

A right triangle with $\overline{AB}$ as the hypothenuse will have, at its intermediate position in Fig.~\ref{fig:subfig2}, the following opposite side to angle $\gamma$:

\begin{equation}
\overline{BC} \cdot\; n\; +\; \overline{OA} \cdot\; m
\label{eq3}
\end{equation}

And the adjacent side:

\begin{equation}
\overline{OC}\; -\; \overline{OA} \cdot\; l\; -\; \overline{BC} \cdot\; o
\label{eq4}
\end{equation}

Applying the Pitagorean theorem in Fig.~\ref{fig:subfig1}:

\begin{equation}
\overline{AB}\; =\; \sqrt{\left(\overline{OC}\; -\; \overline{BC}\right)^{2}\; +\; \left(\overline{OA}\right)^{2}}
\label{eq5}
\end{equation}
    
These relations are needed to obtain $\beta$, $\gamma$, and the equations for velocities and accelerations in the following sections.


\subsection{Velocities}

With pin A viewed as belonging to the disk, its velocity is:

\begin{equation}
\vec{V}_A =  \vec{\omega} \times \vec{\rho}_{OA}
\label{eq6}
\end{equation}

Where $\vec{\omega}\ =\ 2\; \hat{k}\ rad/s$ for the constant rotation case, or $\vec{\omega}\ =\ \left(~0.10~+~2.3\cdot\,\alpha \right) \hat{k}\ rad/s$ for the variable rotation one. For $\alpha$ going from $0^o$ to $90^o$ counterclockwise starting at the lowest position, $\omega$ varies from $0.10\ rad/s$ to $3.7\ rad/s$ when $\omega$ varies linearly with $\alpha$. This accounts for a $1.9\ rad/s$ average angular velocity. We can thus write:

\begin{equation}
  \alpha(t)\ =\ 1.9\cdot\,t
\label{eq6a}
\end{equation}

Taking the time derivative of $\omega$ we get:

\begin{equation}
\dot{\omega}\ =\ 2.3\cdot\,\dot{\alpha}
\label{eq6b}
\end{equation}

And

\begin{equation}
  \dot{\omega}\ =\ \left(2.3 \right)\,\left(1.9 \right)\ =\ 4.4\,rad/s^2
\label{eq6c}
\end{equation}

With pin B belonging to the output bar BC, its velocity is:

\begin{equation}
\vec{V}_B = \vec{\omega}_{BC} \times \vec{\rho}_{CB}
\label{eq7}
\end{equation}

Where $\vec{\omega}_{BC}$ is the still unknown angular velocity vector of bar BC. The motion of linking bar AB can be described using Chasles theorem~\citep{Shames:1996}:

\begin{equation}
\vec{V}_A = \vec{V}_B + \vec{\omega}_{AB} \times \vec{\rho}_{BA}
\label{eq8}
\end{equation}

Where $\vec{\omega}_{AB}$ is the still to be determined angular velocity vector. $\vec{V}_A$ and $\vec{V}_B$ are always tangential to the circular trajectories of pins A and B. For the mechanism's intermediate position shown in Fig.~\ref{fig:subfig2}, $x$ and $y$ components for both velocities can be written as:

\begin{align}
      V_{Ax}&=m \cdot V_A&\ \ \ \ V_{Ay}&=l \cdot V_A \label{eq8a} \\
      V_{Bx}&=n \cdot V_B&\ \ \ \ V_{By}&=o \cdot V_B \label{eq8b}
\end{align}

The fixed vectors in the rigid bodies going from O to A $\left(\vec{\rho}_{OA}\right)$, C to B $\left(\vec{\rho}_{CB}\right)$, and B to A $\left(\vec{\rho}_{BA}\right)$ can also be expressed using their $x$ and $y$ components. An inspection of Fig.~\ref{fig:subfig2} allows one to write:

\begin{align}
      \rho_{OAx}&= + l \cdot \overline{OA} &\ \ \ \ \rho_{OAy}&= - m \cdot \overline{OA} \label{eq8c} \\
      \rho_{CBx}&= - o \cdot \overline{BC} &\ \ \ \ \rho_{CBy}&= + n \cdot \overline{BC} \label{eq8d} \\
      \rho_{BAx}&= - q \cdot \overline{AB} &\ \ \ \ \rho_{BAy}&= - p \cdot \overline{AB} \label{eq8e}
\end{align}

Substituting (\ref{eq8a}) -- (\ref{eq8e}) back in Eqs.~(\ref{eq6}),~(\ref{eq7}), and~(\ref{eq8}):

\begin{equation}
V_B = V_A \cdot \frac{m \cdot q + l \cdot p}{n \cdot q + o \cdot p}
\label{eq9}
\end{equation}

and

\begin{equation}
\omega_{AB} = \frac{V_A}{\overline{AB}}\cdot \left(\frac{m \cdot o - l \cdot n}{o \cdot p + n \cdot q}\right)
\label{eq10}
\end{equation}

With~(\ref{eq9}) in Eqs.~(\ref{eq7}) and~(\ref{eq8b}), $\omega_{BC}$ can be calculated.


\subsection{Accelerations}

We now need to distinguish between the \emph{constant} and the \emph{variable} $\omega$ scenarios.
Pin A only has radial acceleration with constant $\omega$, but it has both radial and tangential components otherwise. Pin B, as before, will have both radial and tangential components.

\subsubsection{Constant $\omega$}
Taking the derivative of Eq.~(\ref{eq8}) with respect to time:

\begin{equation}
\vec{a}_A = \vec{a}_{BR} + \vec{a}_{BT} + \dot{\vec{\omega}}_{AB} \times \vec{\rho}_{BA} + \vec{\omega}_{AB} \times \left( \vec{\omega}_{AB} \times \vec{\rho}_{BA} \right)
\label{eq11}
\end{equation}

The linear accelerations $\vec{a}_A$, $\vec{a}_{BR}$ and $\vec{a}_{BT}$ can be expressed in terms of unit vectors $\hat{i}$ and $\hat{j}$:

\begin{equation}
\vec{a}_A = \left( \omega^{2}\, \overline{OA} \right) \cdot \left(-l\, \hat{i} + m\, \hat{j} \right)
\label{eq12}
\end{equation}

\begin{equation}
\vec{a}_{BR} = \left( {\omega_{BC}}^{2}\, \overline{BC} \right) \cdot \left(o\, \hat{i} - n\, \hat{j} \right)
\label{eq13}
\end{equation}

\begin{equation}
\vec{a}_{BT} = - \left( \dot{\omega}_{BC} \overline{BC} \right) \cdot \left(n\, \hat{i} + o\, \hat{j} \right)
\label{eq14}
\end{equation}

Substituting back into Eq.~(\ref{eq11}) and working the vector algebra, one is able to obtain expressions for angular accelerations of bars AB and BC. As these expressions are long, they are grouped into four different terms as $AB_1$ through $AB_4$ for $\dot{\omega}_{AB}$, and $BC_1$ through $BC_4$ for $\dot{\omega}_{BC}$. It should be noted that these quantities point along the z-direction. The final result is:

\begin{align}
      AB_{1}&=\omega^2\;\overline{OA}\cdot \left( l o + m n \right)&\ \ \ \ AB_{2}&={\omega_{BC}}^2\;\overline{BC}\cdot \left(o^2 + n^2 \right) \label{eq15a}\\
      AB_{3}&={\omega_{AB}}^2\;\overline{AB}\cdot \left( o q - n p \right)&\ \ \ \ AB_{4}&=\overline{AB}\;\cdot \left(o p + n q \right) \label{eq15b}
\end{align}

\begin{equation}
\dot{\omega}_{AB} = - \frac{AB_1 + AB_2 + AB_3}{AB_4}
\label{eq15}
\end{equation}

\begin{align}
      BC_{1}&=\omega^2\;\overline{OA}\cdot \left( m p - l q \right)&\ \ \ \ BC_{2}&={\omega_{AB}}^2\ \overline{AB} \cdot \left(p^2 + q^2 \right) \\
      BC_{3}&={\omega_{BC}}^2\;\overline{BC}\cdot \left( n p - o q  \right)&\ \ \ \ BC_{4}&=\overline{BC}\;\cdot \left(n q + o p \right)
\end{align}

\begin{equation}
\dot{\omega}_{BC} = - \frac{BC_1 - BC_2 + BC_3}{BC_4}
\label{eq16}
\end{equation}

\subsubsection{Variable $\omega$}

Since the disk now has a variable rate of rotation $\dot{\omega}$, pin A's acceleration is:

\begin{equation}
\vec{a}_A = \vec{a}_{AR}~+~\vec{a}_{AT}
\label{eq17}
\end{equation}

Referring back to Fig.~\ref{fig:subfig2}, the tangential component can be written as:

\begin{equation}
  \vec{a}_{AT}~=~\dot{\omega}\,\overline{OA}\,\left( m \hat{i}\,+\,l \hat{j} \right)
\end{equation}

The radial component is still given by Eq.~(\ref{eq12}). As for the constant case, angular accelerations for bars AB and BC are written as:

\begin{align}
      AB_{1}&=\omega^2\;\overline{OA}\cdot \left( l o + m n \right)&\ \ \ \ AB_{2}&={\dot{\omega}}\;\overline{OA}\cdot \left(m o - l n \right) \label{eq18a}\\
      AB_{3}&={\omega_{BC}}^2\;\overline{BC}\cdot \left(o^2 + n^2 \right)&\ \ \ \ AB_{4}&={\omega_{AB}}^2\;\overline{AB}\cdot \left( o q - n p \right) \label{eq18b}\\
      AB_{5}&=\overline{AB}\;\cdot \left(o p + n q \right)&\ \ \ \ \label{eq18c}\\
\end{align}

\begin{equation}
\dot{\omega}_{AB} = \frac{-AB_1 + AB_2 - AB_3 - AB_4}{AB_5}
\label{eq18}
\end{equation}

And,

\begin{align}
      BC_{1}&=\omega^2\;\overline{OA}\cdot \left( m p - l q \right)&\ \ \ \ BC_{2}&={\dot{\omega}}^2\ \overline{OA} \cdot \left(m q + l p \right) \\
      BC_{3}&={\omega_{BC}}^2 \;\overline{BC}\cdot \left( n p - o q  \right)&\ \ \ \ BC_{4}&={\omega_{AB}}^2\ \overline{AB} \cdot \left(p^2 + q^2 \right) \\
      BC_{5}&=\overline{BC}\cdot \left( n q + o p  \right)&\ \ \ \ 
\end{align}

\begin{equation}
\dot{\omega}_{BC} = \frac{-BC_1 - BC_2 - BC_3 + BC_4}{BC_5}
\label{eq19}
\end{equation}

This concludes the kinematical formulation of the articulated mechanism that has been implemented numerically and placed under \emph{git} version control~\citep{Fleischfresser:2014}.

\section{NUMERICAL EXPERIMENTS}

\subsection{Constant $\omega$}
\begin{figure*}[t]
\centering
\includegraphics[angle=0, scale=0.45]{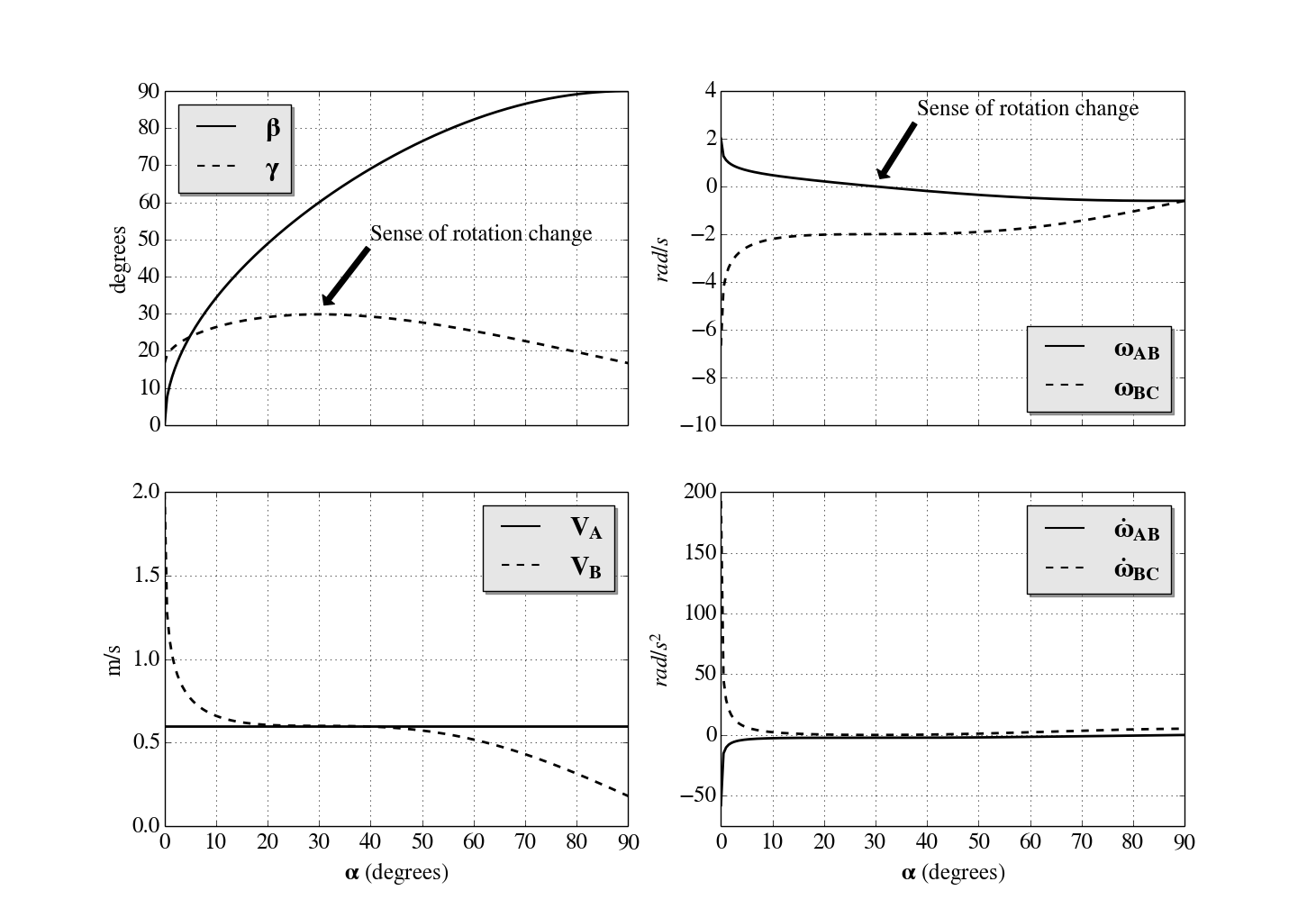}
\caption{\textit{Top left}: $\beta$ and $\gamma$ vs. $\alpha$. \textit{Bottom left}: linear velocities of pins A and B vs. $\alpha$. \textit{Top right}: angular velocities of bars AB and BC vs. $\alpha$. \textit{Bottom right}: angular accelerations of bars AB and BC vs. $\alpha$. Angular step size $\Delta \alpha\,=\,0.5^{o}$.}
\label{fig4}
\end{figure*}

Geometrical and kinematics parameters for the $90^{o}$ turn of the disk are shown in Fig.~\ref{fig4}. Of note is the change of rotation sense for the linking bar AB from counterclockwise (positive $\omega_{AB}$) to clockwise (negative $\omega_{AB}$) when $\alpha = 30^{o}$ (top right panel). This feature is also evident when $\gamma$ reaches a maximum (top left panel), and $V_B$ and $\omega_{BC}$ remain nearly constant around $\alpha = 30^{o}$ (bottom left and top right panels). Angular accelerations, despite their intricate mathematical expressions [Eqs.~(\ref{eq15a}) through~(\ref{eq16})], are slowly varying functions after the start-up period. 

\begin{table}[!ht]
\begin{center}
\caption{Head and tail of linear and angular velocities. See text for comments on cells with asterisks.}
\begin{tabular}{|c|c|c|c|c|c|c|c|}
\hline
$\alpha$ ($^{o}$) & $\beta$ ($^{o}$) & $\gamma$ ($^{o}$) & $V_A$ ($m/s$) & $V_B$ ($m/s$) & $\omega$ ($rad/s$) & $\omega_{AB}$ ($rad/s$) & $\omega_{BC}$ ($rad/s$) \\
\hline
 0 & 0 & 16.7 & 0.6 &  2* &  2* &  2* & -6.7\\
 \hline
 \vdots & \vdots & \vdots & \vdots & \vdots & \vdots & \vdots & \vdots \\
\hline
 90 & 90 & 16.7 &  0.6* & 0.18 & 2 & -0.6* & -0.6* \\
\hline
\end{tabular}
\label{tab1}
\end{center}
\end{table}

\pagebreak

Table~\ref{tab2} has the computed angular accelerations' initial and final values. The results for $\alpha\,=\,0^{o}$ serve to validate the numerical implementation since the computations match the textbook calculations in~\cite{Shames:1996}. The same is true for the angular velocities of Tab.~\ref{tab1}.

\begin{table}[!ht]
\begin{center}
\caption{Head and tail of angular accelerations.}
\begin{tabular}{|c|c|c|c|c|}
\hline
$\alpha$ ($^{o}$) & $\beta$ ($^{o}$) & $\gamma$ ($^{o}$) & $\dot{\omega}_{AB}$ ($rad/s^2$) & $\dot{\omega}_{BC}$ ($rad/s^2$)\\
\hline
 0 & 0 & 16.7 & -57.8 & 192.6\\
 \hline
 \vdots & \vdots & \vdots & \vdots & \vdots \\
\hline
 90 & 90 & 16.7 & 0.0 & 5.2 \\
\hline
\end{tabular}
\label{tab2}
\end{center}
\end{table}

\subsection{Variable $\omega$}

\begin{figure*}[h]
\centering
\includegraphics[angle=0, scale=0.45]{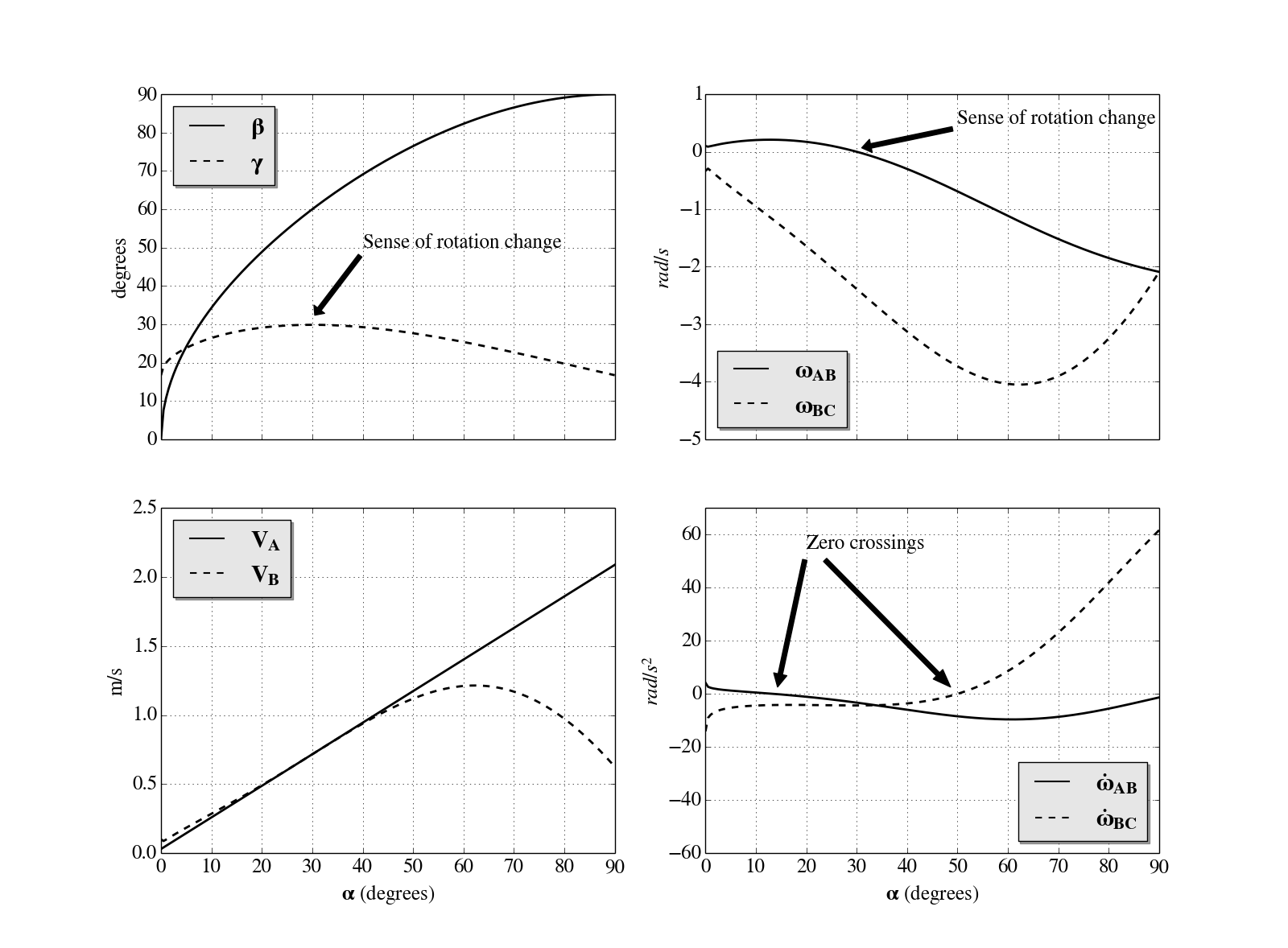}
\caption{\textit{Top left}: $\beta$ and $\gamma$ vs. $\alpha$. \textit{Bottom left}: linear velocities of pins A and B vs. $\alpha$. \textit{Top right}: angular velocities of bars AB and BC vs. $\alpha$. \textit{Bottom right}: angular accelerations of bars AB and BC vs. $\alpha$. Angular step size $\Delta \alpha\,=\,0.5^{o}$.}
\label{fig5}
\end{figure*}

Results are now displayed in Fig.~\ref{fig5}. Note the linear increase of $V_A$ (bottom left panel), and how angular speeds and accelerations evolve without resemblance to the constant $\omega$ case. In particular, it is now easy to see when both bars start decelerating during the $90^{o}$ counterclockwise turn.

\begin{table}[!ht]
\begin{center}
\caption{Head and tail of linear and angular velocities. See text for comments on cells with asterisks.}
\begin{tabular}{|c|c|c|c|c|c|c|c|}
\hline
$\alpha$ ($^{o}$) & $\beta$ ($^{o}$) & $\gamma$ ($^{o}$) & $V_A$ ($m/s$) & $V_B$ ($m/s$) & $\omega$ ($rad/s$) & $\omega_{AB}$ ($rad/s$) & $\omega_{BC}$ ($rad/s$) \\
\hline
 0 & 0 & 16.7 & 0.03 &  0.1* &  0.1* &  0.1* & -0.33\\
 \hline
 \vdots & \vdots & \vdots & \vdots & \vdots & \vdots & \vdots & \vdots \\
\hline
 90 & 90 & 16.7 &  2.1* & 0.63 & 7.0 & -2.1* & -2.1* \\
\hline
\end{tabular}
\label{tab3}
\end{center}
\end{table}

\begin{table}[!ht]
\begin{center}
\caption{Head and tail of angular accelerations.}
\begin{tabular}{|c|c|c|c|c|}
\hline
$\alpha$ ($^{o}$) & $\beta$ ($^{o}$) & $\gamma$ ($^{o}$) & $\dot{\omega}_{AB}$ ($rad/s^2$) & $\dot{\omega}_{BC}$ ($rad/s^2$)\\
\hline
 0 & 0 & 16.7 & 4.2 & -14.1\\
 \hline
 \vdots & \vdots & \vdots & \vdots & \vdots \\
\hline
 90 & 90 & 16.7 & -1.3 & 61.7 \\
\hline
\end{tabular}
\label{tab4}
\end{center}
\end{table}

\section{DISCUSSION AND CONCLUSIONS}
The relations needed to simulate the motion of an articulated mechanism were presented and the kinematics was analyzed numerically. Results for initial and final positions ($\alpha = 0^{o}$ and $90^{o}$ respectively) are shown in Tab.~\ref{tab1}, Tab.~\ref{tab2}, Tab.~\ref{tab3} and Tab.~\ref{tab4}. At the initial position ($\alpha = 0^{o}$) both $V_B$ and $\omega_{AB}$ have the same numerical values of the disk's rotation rate (cells with asterisks). Paper and pencil solutions give instantaneous values while numerical experiments capture the motion's evolution. Hence, if only a paper and pencil solution is attempted for $\alpha = 0^{o}$, one may be tempted to believe these parameters always have the same numerical values. The same would be true if one attempts a paper and pencil solution for the final position ($\alpha = 90^{o}$), since $\omega_{AB}$ and $\omega_{BC}$ have the same absolute values of the constant linear speed of a point at the edge of the disk (cells with asterisks).

\begin{figure*}[h]
\centering
\includegraphics[angle=0, scale=0.7]{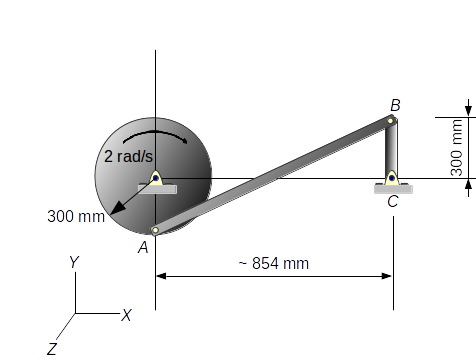}
\caption{Articulated mechanism design modification that allows for full disk revolution.}
\label{fig6}
\end{figure*}

The experiments also demonstrated the inversion of rotation for the linking bar under plane motion for two different scenarios: one where the rate of rotation of the disk is constant, and one where it is made to vary linearly with the angle of rotation. Analysis such as the one presented here complement routine paper and pencil solutions which alone may lead to erroneous conclusions about the motions. A design modification of the original mechanism is shown in Fig.~\ref{fig6}. The distance between the fixed centers of rotation is shorter than the original mechanism to allow full disk revolution while keeping the linking bar size unchanged, and the disk rotation is now clockwise. The full revolution is not possible in the original configuration (Fig.~\ref{fig:both}).

Despite the complexities of the rigid bodies' formulations described in this article, one should keep in mind that this is still a relatively simple mechanics problem. All principles applied are exact and the numerical implementations only required algebraic expressions. Complications arise if one attempts to formulate and model collisions between rigid bodies. Impulse-momentum principles are needed compared to the limited kinematics toolbox. Moreover, collision laws are not well understood and numerical implementations need to handle differential equations with discontinuities that may lead to energy conservation inconsistencies and other difficulties~(\citep{Stewart:2000},~\citep{Garstenauer:2006},~\citep{Chatterjee:1997}).

\newpage

\begin{thebibliography}{9}   


\bibitem{Stewart:2000}
Stewart,~D.~E.~(2000).
\newblock Rigid-Body Dynamics with Friction and Impact.
\newblock {\em SIAM Review}, 42:3--39.

\bibitem{Garstenauer:2006}
Garstenauer,~H.~(2006).
\newblock A Unified Framework for Rigid Body Dynamics.
\newblock {\em Ph.D.~Thesis.}
\newblock {Johannes Kepler Universit\"{a}t.}
\newblock {Linz.}

\bibitem{Shames:1996}
Shames,~I.~H.~(1996).
\newblock Engineering Mechanics: Statics and Dynamics.
\newblock {\em Prentice Hall.}
\newblock {ISBN: 0-13-356924-1}

\bibitem{Fleischfresser:2014}
Fleischfresser,~L.~(2014).
\newblock Scaling-octo-tribble: Kinematics of disk-bars mechanism.
\newblock {DOI: 10.5281/zenodo.12417.}
\newblock {\em Available at:}
\newblock {https://dx.doi.org/10.5281/zenodo.12417}

\bibitem{Chatterjee:1997}
Chatterjee,~A.~(1997).
\newblock Rigid Body Collisions: Some General Considerations, New Collision Laws, and Some Experimental Data.
\newblock {\em Ph.D.~Thesis.}
\newblock {Cornell University.}
\newblock {Ithaca.}





\end{thebibliography}
\end{document}